%% file: p.tex
\documentclass[times, authoryear]{elsarticle}

\usepackage{color,graphicx}
\usepackage{mathtools}
\usepackage{amsfonts}
\usepackage{hyperref}

\newcommand{\vect}[1]{\mathbf {#1} }

\begin{document}

\begin{frontmatter}
\title{Importance of the numerical schemes \\ in the CFD of the human nose}
\author{A. Schillaci}
\ead{andrea.schillaci@polimi.it}
\author{M. Quadrio \corref{cor1}}
\ead{maurizio.quadrio@polimi.it}
\cortext[cor1]{Corresponding author}
\address{Department of Aerospace Science and Technology \\
Politecnico di Milano \\
Via La Masa, 34
20156 Milano - Italy }

\begin{abstract}
Computational fluid dynamics of the air flow in the human nasal cavities, starting from patient-specific Computer Tomography (CT) scans, is an important tool for diagnostics and surgery planning. However, a complete and systematic assessment of the influence of the main modeling assumptions is still lacking. In designing such simulations, choosing the discretization scheme, which is the main subject of the present work, is an often overlooked decision of primary importance. We use a comparison framework to quantify the effects of the major design choices on the results. The reconstructed airways of a healthy, representative adult patient are used to set up a computational study where such effects are systematically measured. It is found that the choice of the numerical scheme is the most important aspect, although all varied parameters impact the solution noticeably. For a physiologically meaningful flow rate, changes of the global pressure drop up to more than 50\% are observed; locally, velocity differences can become extremely significant. Our results call for an improved standard in the description of this type of numerical studies, where way too often the order of accuracy of the numerical scheme is not mentioned.
\end{abstract}

\end{frontmatter}

\input{intro}      
\input{methods}    
\input{results}    
\input{discussion} 

\section{Conclusion}
The impact of key methodological choices in the numerical simulation of the airflow in the human nasal cavities has been quantitatively assessed, by comparing the importance of the numerical scheme accuracy to that of the flow modelling. Within a well-defined comparison framework, the output of 24 simulations has been evaluated at both the global and local level in terms of pressure losses, mean velocity and pressure fields. The choice of a laminar/RANS/LES modelling approach is very important, especially in such flows that are often laminar, albeit vortical, chaotic and three-dimensional. However, we have ascertained that the numerical scheme is even more important, leading to differences to more than 50\% in global indicators (e.g. nasal resistance), and to local differences that can be extremely significant. Finally, we have also indirectly assessed that cone-beam CT scans can be used proficiently, at long as inspiration is considered; in expiration, however, the proximity of the inflow to the nasopharynx is responsible for a significant misrepresentation of the laryngeal jet that propagates up to the nostrils. Overall, the study confirms that high-fidelity and time-resolved LES/DNS computations \citep{calmet-etal-2020} are probably necessary for a reliable simulation of the full breathing cycle at intermediate intensity, and advocates once again for high-quality numerical and experimental benchmarks, placed on the public domain and fully reproducible, to arrive at a rigorous assessment of the adequacy of the modelling choices in the CFD of the nasal airflow.

\section*{Conflict of interest statement}
The authors declare that they have no competing financial interests or personal relationships that could have influenced the work reported in this paper.

\section*{Acknowledgments}
Computing time was provided by the CINECA Italian Supercomputing Center.

\bibliographystyle{elsarticle-harv}

\section*{Appendix}

\section*{A1. The numerical approach}

\subsection*{A1.1 Boundary conditions}

Regardless of the flow modelling approach, at the boundary made by solid, rigid walls the velocity vector and the normal component of the pressure gradient are set to zero. During inspiration, the inlet is the surface of the external sphere surrounding the nose tip, and the outlet is at the throat. The required flow rate of 280 $ml/s$ is prescribed at the inlet via an inward velocity field that is computed to be locally normal to the surface, and adjusted to provide the prescribed integral value. This value of the flow rate is representative of slow to mild breathing \citep{wang-lee-gordon-2012,covello-etal-2018}. Pressure at the inlet is given a zero-gradient condition. At the outlet, the gradient of the velocity is set to zero, and the total pressure is set to a given (zero) reference value. During expiration, the boundary conditions are reversed: the throat becomes the inlet, where the flow rate of 280 $ml/s$ is prescribed via the normal velocity field adjusted to provide the specified integral, and pressure has zero gradient. At the outlet, i.e. the external sphere, a zero-gradient condition for velocity is accompanied by a reference zero value for the total pressure.

\subsection*{A1.2 RANS model and procedures}

The model of choice, already used in the past for such studies \citep[see e.g.][]{liu-etal-2007,li-etal-2017}, is the $k$--$\omega$--SST model: besides the RANS equations, it solves two additional partial differential equations for the turbulent kinetic energy $k$ and the turbulent frequency $\omega$, and uses the Bousinnesq hypothesis to close the RANS equations via a turbulent viscosity $\nu_t$. The turbulent frequency at the wall provides a wall constraint on the specific dissipation rate, and is imposed to match the condition by \cite{menter-kuntz-langtry-2003}. At the inlet, the flow is considered nearly non-turbulent, and the turbulent frequency is thus set to an arbitrarily small value (unitary in the present work). At the outlet its gradient is null. The turbulent kinetic energy at the wall is zero by definition. At the inlet it is set to $k = \frac{1}{2} (I \mathbf{U})^2$ where $\mathbf{U}$ is the local (extremely small) mean velocity and $I$ is the turbulent intensity, which is set to 2\%, resulting in an almost non-turbulent inlet. At the outlet the gradient of $k$ is zero. The $k$--$\omega$--SST model is known for its ability to provide results that are decoupled from the (necessarily approximate) far-field values of the boundary conditions.

The model coefficients are standard and taken from \cite{menter-kuntz-langtry-2003}: 

\begin{align*}
\sigma_{k_1} &= 0.85;       & \sigma_{k_2} &= 1.0;          & \sigma_{\omega_1} &= 0.5; \\
\sigma_{\omega_2} &= 0.856;  & \alpha_1 &= \frac{5}{9};  & \alpha_2 &= 0.44;\\
\beta_1 &= \frac{3}{40};          & \beta_2 &= 0.0828;        & \beta^* &= 0.09.
\end{align*}

The iterative solution of the RANS equations is terminated when the residuals fall below set limits. The typical limit values for RANS-I are $10^{-9}$ for the residuals of the three components of velocity and pressure, and $10^{-6}$ for the turbulent quantities. These values are set higher for RANS-II and become $10^{-5}$ for every quantity. This level is considered sufficient to yield fully converged results \citep{zhang-etal-2011}.

\subsection*{A1.3 LES model and procedures}

The model of choice is the Wall-Adapting Local Eddy-viscosity (WALE) model, already employed in similar studies \citep{li-etal-2017}. It is an algebraic (hence not requiring additional boundary conditions) sub-grid scale model based on the work by \cite{nicoud-ducros-1999}. The model considers both local strain and rotation rates, and has the advantage of being invariant to translation and rotation of the reference system; moreover, it requires local information only, which makes it particularly useful for complex geometries such as the present one. Compared to the widely used Smagorinsky model \citep{smagorinsky-1963}, by design WALE provides an eddy viscosity that approaches zero at a solid wall with the correct rate. 

The time-dependent solution of the LES equations is stopped when the length of the statistical sample is enough to provide satisfactory estimates for the first and second statistical moments of the flow variables. We know by previous experience \citep{covello-etal-2018} that, at this respiratory rate, a time integration of 0.6 seconds after properly discarding the initial transient is enough to provide a reliable mean field. In fact, a test carried out for CT-LES-II-i by adding 0.20 $s$ of simulation time has led to change in the total pressure drop by 0.017\% only.
 
The time-dependent LES cases are advanced in time with a second-order BDF (backward differentiation formula) scheme; the time step is dynamically varied so that the Courant--Friedrichs--Lewy number remains below one. For example, for CT-LES-II-i the average value of the timestep is $1.038 \times 10^{-5}$ seconds, with maximum and minimum during the simulation being $1.226 \times  10^{-5}$ and $7.809 \times  10^{-6}$ seconds. The obtained average timestep is in line with the values used in literature \citep{li-etal-2017} for comparable LES simulations.

\begin{figure}
\centering
\includegraphics[width=0.8\columnwidth]{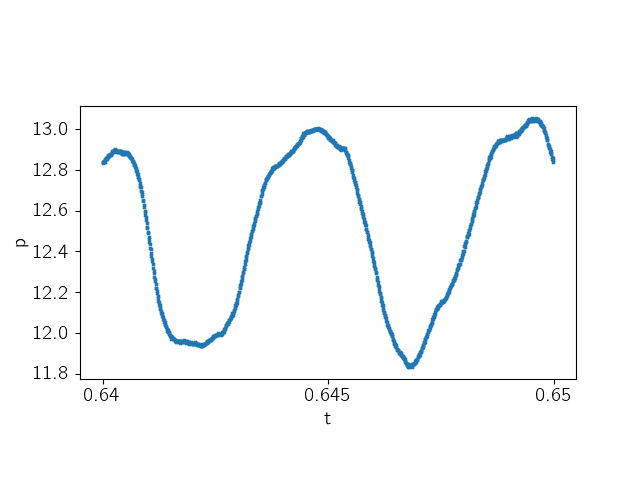}
\caption{Time history (case CT-LES-II-i) over 1200 time steps of the instantaneous pressure signal measured by a probe positioned in the rhinopharynx in the point with coordinates $x=0.001$, $y=0.05$ and $z=0.0145$. Each dot represents one time step.}
\label{fig:probe}
\end{figure}

A reliable estimate of the frequency content of the temporal signal cannot be obtained with classical turbulence arguments (e.g. comparison with the local Kolmogorov time scale), owing to the significant difference between this flow and the homogeneous isotropic turbulent flow where such arguments apply. Although the only proper assessment would derive from a frequency analysis of the temporal signal at every spatial location (a non-trivial endeavour for a non-periodic signal known over a finite time horizon), visual inspection of the time history of the local turbulent signal would provide a good proxy to establish whether the time step is small enough. Figure \ref{fig:probe} plots the time history of pressure measured at one spatial location in the rhinopharynx, where most unsteadiness takes place. One immediately obtains the visual evidence that the temporal signal is extremely well sampled, reinforcing the concept \citep{choi-moin-1994,bernardini-etal-2013} that in DNS (or high-resolution LES) of wall-bounded flows the time step is usually dictated by the stability margin of the numerical scheme, well below the temporal scales of the turbulent flow.

\section*{A2. Geometry and mesh quality}

The CT scan is obtained from an adult male with healthy sino-nasal anatomy. The scan has an axial resolution of $0.6 mm$, with coronal and sagittal spacing of $0.5 mm$, and is representative of current clinical standards. The CT scan extends down to the larynx and includes the whole volume of the sinuses. The image set from the CT scan is processed via 3D-Slicer \citep{fedorov-etal-2012}, an open-source software for the analysis and visualization of medical images. Without any other manipulation or smoothing, 3D-Slicer is used, via image segmentation at uniform threshold of $-220$ Hounsfield units and volume reconstruction, to convert the CT images into a three-dimensional geometry, that is eventually exported as a STereoLithography (STL) file. 

All the computational volume meshes are generated within the OpenFOAM toolbox \citep{weller-etal-1998}: in particular the \textit{snappyHexMesh} tool is used to  convert the STL geometry into a computational mesh. First a volume surrounding the natomy is filled with hexahedra, then a refinement is applied to the surface of the geometry such that the cells are decomposed twice in each direction (castellation phase). In this way, a background cells that intersect the STL surface is split in 64 smaller cells. The vertices of the castellated mesh are then displaced to snap onto the actual boundary (snap phase). No additional prismatic layers are added, to avoid deterioration of the mesh quality. The mesh is fine enough at the wall, with mean values of about $y^+ = 0.5$ for LES simulations and of around $y^+ = 1.09$ for the laminar/RANS simulations (with $y$ denoting, as it is customary, the wall-normal distance, and the plus superscript indicating viscous or wall units). The mesh refinement near the wall can be seen in Fig. 1b of the main manuscript. The dictionary controlling the mesh quality is the same for LES and laminar/RANS meshes: the finer LES meshes are obtained by starting from a finer background mesh. 

In terms of mesh quality parameters, for the RANS simulation (total cells about 3.2 millions), the average non-orthogonality is 11.86, well below the limit margin of 70, above which special treatment would be required. The maximum skewness is 2.46, which is above the warning threshold of 0.9 but fully acceptable in a geometry of this complexity. 

\begin{figure}
\centering
\includegraphics[width=\textwidth]{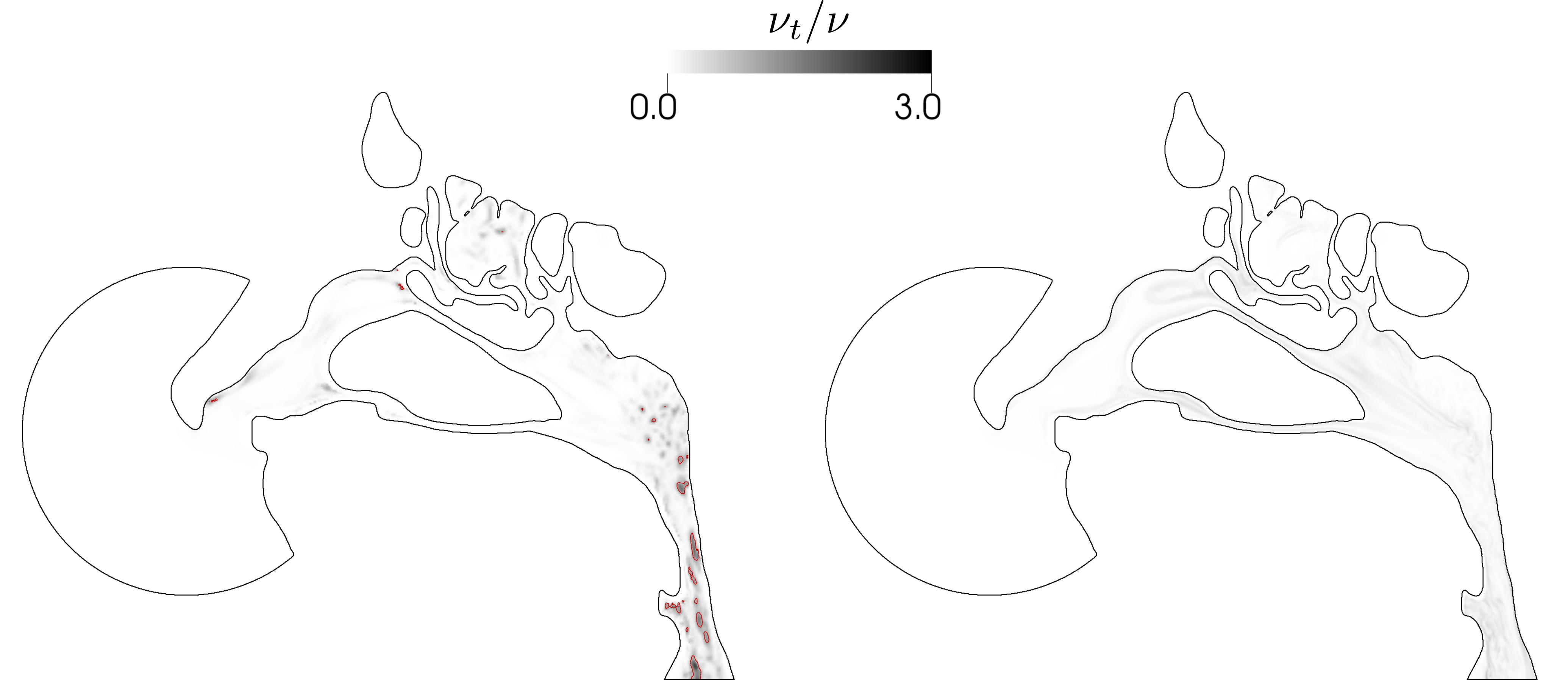}
\caption{Ratio between turbulent viscosity $\nu_t$ and molecular viscosity $\nu$ for an instantaneous flow field of the case CT-LES-II-i at 15 million cells (left) and HRLES-II-i at 50 million cells (right). The red contour line marks the unitary ratio.}
\label{fig:nut_LES}
\end{figure}

For the LES mesh (total cells about 15 millions) the average non-orthogonality is 11.71, and the maximum cell skewness is 2.49, i.e. nearly identical to the RANS mesh. This is because the mesh is generated with the same target quality parameters. Fig.\ref{fig:nut_LES} shows that the LES mesh is quite fine for the present problem, and yields a ratio between the turbulent viscosity $\nu_t$ and the molecular viscosity $\nu$ that, in one randomly chosen instantaneous field, remains below 2.5. Thus the LES is certainly well resolved, especially near the wall, with the model yielding a limited contribution. (The same ratio remains well below unity everywhere for the larger HRLES mesh, which could thus be considered a fully resolved DNS.)

\begin{figure}
\centering
\includegraphics[width=0.6\columnwidth]{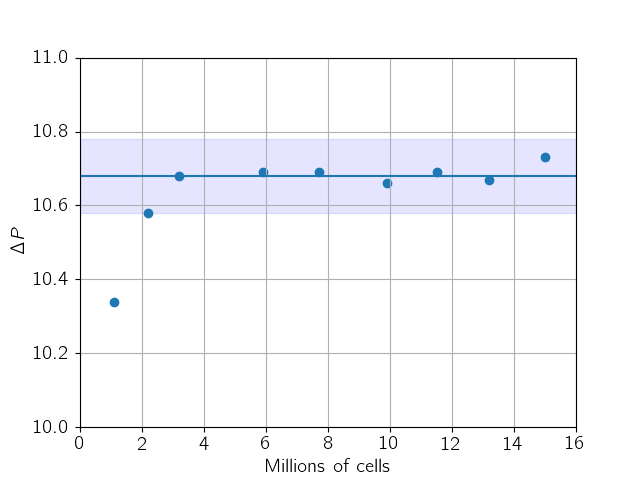}
\caption{Grid convergence study for the case CT-RANS-II-i. Pressure drop (measured between external ambient and the lower end indicated by the red line in figure 1a of the main text) is plotted versus mesh size. The horizontal band shows the mean value (excluding the first two leftmost points) of the pressure drop, and the $\pm 1\%$ interval.}
\label{fig:grid_convergence}
\end{figure}

While the fineness of the LES mesh has been evaluated above in the context of fig.\ref{fig:nut_LES}, a grid refinement study has been carried out for the RANS mesh, to confirm that its size, which is typical of comparable literature studies employing RANS, is indeed adequate. The refinement study, whose results are shown in figure  \ref{fig:grid_convergence}, proves that the considered mesh at 3.2 millions cells is already fine enough to properly capture the pressure drop between the external ambient and the outlet. Too small a mesh (below two millions cells) is clearly insufficient to describe the geometry properly, but all the finer meshes show that results are well within a $\pm 1\%$ uncertainty band.

\section*{A3. CT vs TrCT}

\begin{figure}
\centering
\includegraphics[width=\columnwidth]{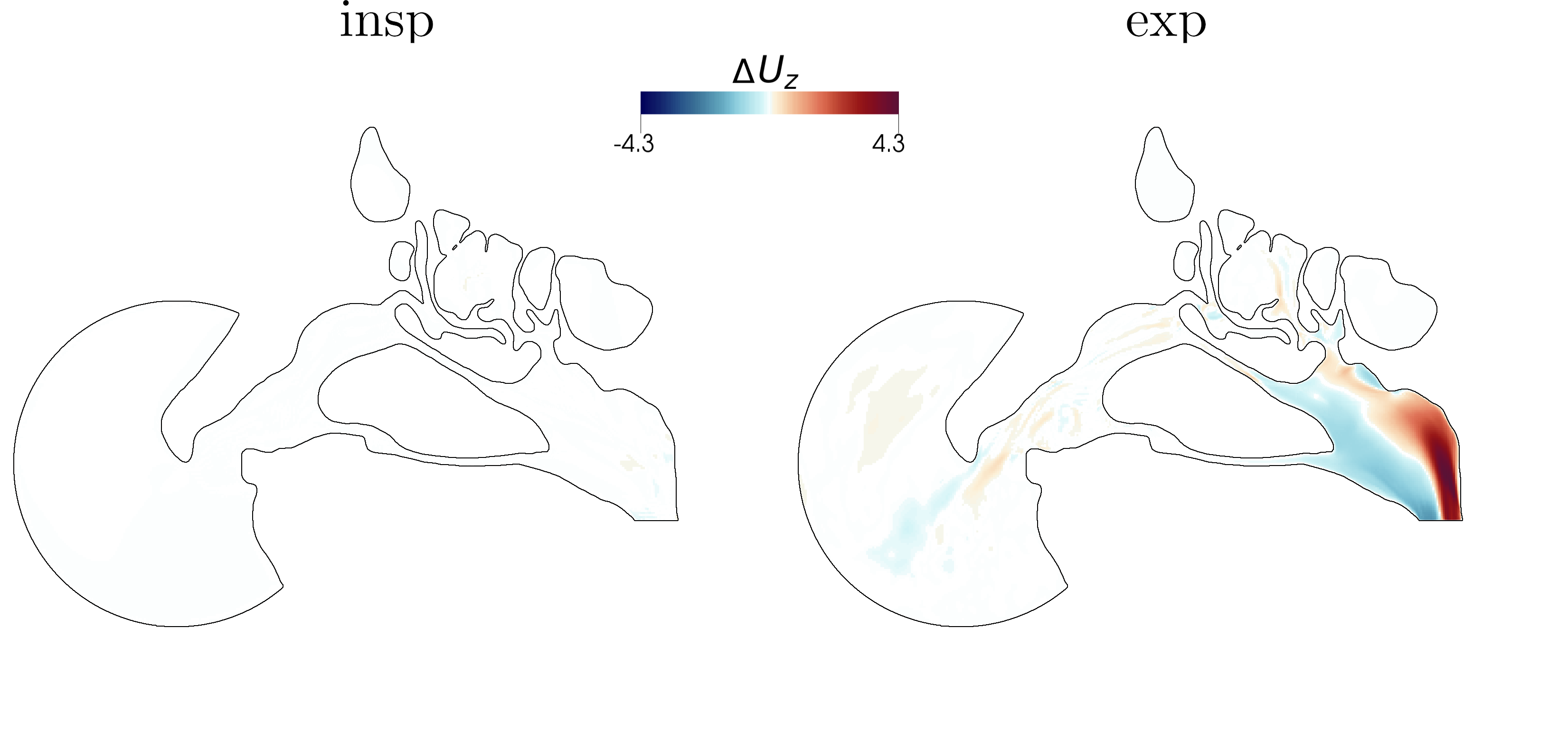}
\caption{Differential velocity field $\vect{U}_{CT} - \vect{U}_{TrCT}$ (vertical component), for LES-II.}
\label{fig:TrCT}
\end{figure}

The effect of the position of the lower boundary of the computational domain is a parameter that in the main text has been discussed only in terms of global changes of the pressure drop. Here we describe in Figure \ref{fig:TrCT} the local differences, by plotting the field $\vect{U}_{CT} - \vect{U}_{TrCT}$. Since this set of differences is mostly insensitive to the turbulence modeling approach, only the LES-II simulations are shown.

The position of the lower boundary shows little or no effect when inspiration is considered, with peak differences as small as 0.2 $m/s$. However, during expiration large differences, of about 4.3 $m/s$, are observed near the inlet, because the structure of the laryngeal jet is not reproduced properly by the smaller-domain simulation. Importantly, these differences decrease with the distance from the inlet but persist quite far from it, and non-negligible effects are discerned even at the nostrils.

\end{document}

%% file: intro.tex
\section{Introduction}
\label{intro}

Nasal breathing difficulties are a widespread pathological condition, accompanied by significant economical and social costs \citep{smith-orlandi-rudmik-2015, rudmik-etal-2015}. A precise diagnosis is often difficult to achieve, corrective surgeries are sometimes required, yet after certain nose surgeries the majority of patients remains unsatisfied \citep{sundh-sonnergren-2015}. 

Starting about two decades ago, numerical studies of nasal airflow based on Computational Fluid Dynamics (CFD) began to increase in number and quality. Nowadays, Ear, Nose and Throat (ENT) doctors envisage the use of a detailed CFD solution to diagnose pathologies and to plan surgeries \citep{radulesco-etal-2020, singh-inthavong-2021}. A recent, broad and insightful account of potential and open problems is given by \cite{inthavong-etal-2019}. 

There is thus a growing need for a thorough validation and standardization of CFD methods and procedures. Several aspects, like the spatial resolution of the computational mesh \citep{frankito-etal-2015b}, or the radio-density threshold employed for CT segmentation \citep{zwicker-etal-2018} have been specifically discussed, but a systematic assessment of the sensitivity of the CFD outcome to the various sources of uncertainty involved in the procedure is still required, noticeably so in respect to the discretization errors incurred by the numerical method. The present work describes and compares within a unified framework two major contributors to the global error in a well conducted CFD simulation: how the flow physics is modeled, and which schemes are used in the numerical solution. The former contribution has been discussed several times, while the latter has never been addressed.

CFD simulations of the nasal airflow nowadays leverage the entire spectrum of flow modeling choices, ranging from Direct Numerical Simulations (DNS) to Large-Eddy Simulations (LES) and Reynolds-averaged Navier--Stokes equations (RANS). Moreover, "laminar" simulations are also employed, where a steady RANS solver is used without a turbulence model under the assumption of steady flow. RANS assumes the flow to be turbulent, employs a (dissipative) turbulence model to describe the effect of the turbulent fluctuating field on the time-averaged motion, and only computes a time-averaged solution via a steady solver; it represents the computationally cheapest approach, with the largest amount of modeling error. DNS is at the other end of the spectrum: it solves the unsteady equations of motion without a turbulence model, because the solution takes place on a spatial mesh fine enough to resolve all the significant flow scales; the obvious downside is the computational cost. LES is midway between the two extrema, but akin to DNS: the solution is time-dependent and relatively expensive from a computational standpoint, while the role of the turbulence model, which is still required, is relatively minor and can be controlled via the size of the mesh. A further option, still used scarcely in this field, is the combined use \citep[see e.g.][]{vanstrien-etal-2021} of RANS and LES with the so called hybrid methods, which are able to bring forth the unsteady character of the flow in the nasopharynx even at low flow rates. 

The importance of flow modelling is well known. For example, Zhao and coworkers \citep{li-etal-2017} thoroughly compared results from several RANS models, one LES model and a reference DNS, for an artificial anatomy deprived of sinuses for which prior experimental information was available. Within a commercial solver, they used second-order numerical schemes for RANS and bounded second-order schemes for LES. The laminar flow model was found to perform well, at low breathing intensity, to predict the pressure drop, but was observed to not excel at predicting local velocity profiles compared to other approaches. In fact, even for steady boundary conditions, the complex anatomy of the nasal cavity may  lead to a three-dimensional and unsteady flow in the nasal fossae of a healthy subject \citep{churchill-etal-2004} which is mostly laminar at low flow rates \citep{chung-etal-2006}, but becomes transitional and/or turbulent at higher respiratory rates, especially in the rhinopharynx. Unsteadiness becomes locally very important, even at slow flow, in presence of anatomic anomalies \citep{saibene-etal-2020}, suggesting LES as the preferred approach, especially when particle tracking is involved \citep{farnoud-etal-2020}. While many valuable contributions \citep{liu-etal-2007,calmet-etal-2020} employ a time-dependent solution, owing to the lower computational cost several works being published nowadays still remain of the laminar or RANS type.

Less attention has been devoted to another important design choice, whose effects are often underestimated, to the point that most papers do not even mention it: one needs to decide how to discretize the differential operators in the equations of fluid motion. In a finite-volumes CFD software (the most widespread approach), it is customary to have at least two choices available, depending on whether differential operators are discretized at first- or second-order accuracy; some codes allow to pick a different scheme for each term in the differential equations. The formal order of accuracy is the integer power of of the cell size that brings the discretization error towards zero \citep{ferziger-peric-2002}.

The present work introduces a comparison framework where the effects of the discretization scheme are quantified and compared to those related to the choice of the flow model  (laminar, RANS or LES/DNS). Additionally, the same framework is used to quantify the effects of a computational domain truncated at the nasopharynx. Studying domain truncation is not new: e.g. \cite{choi-etal-2009} did a similar study for the flow in the lungs, but only considered lower truncations below the larynx with breathing through the mouth. In the present context, and in view of the increasing availability of cone-beam CT scanners, which impart smaller radiation dosages with better spatial resolution at the cost of a smaller field of view \citep{tretiakow-etal-2020}, it is interesting to observe the effects of domain truncation just after the nasal fossae.

%% file: methods.tex
\section{Methods}
\label{methods}

\begin{figure}
\centering
\includegraphics[width=0.6\columnwidth]{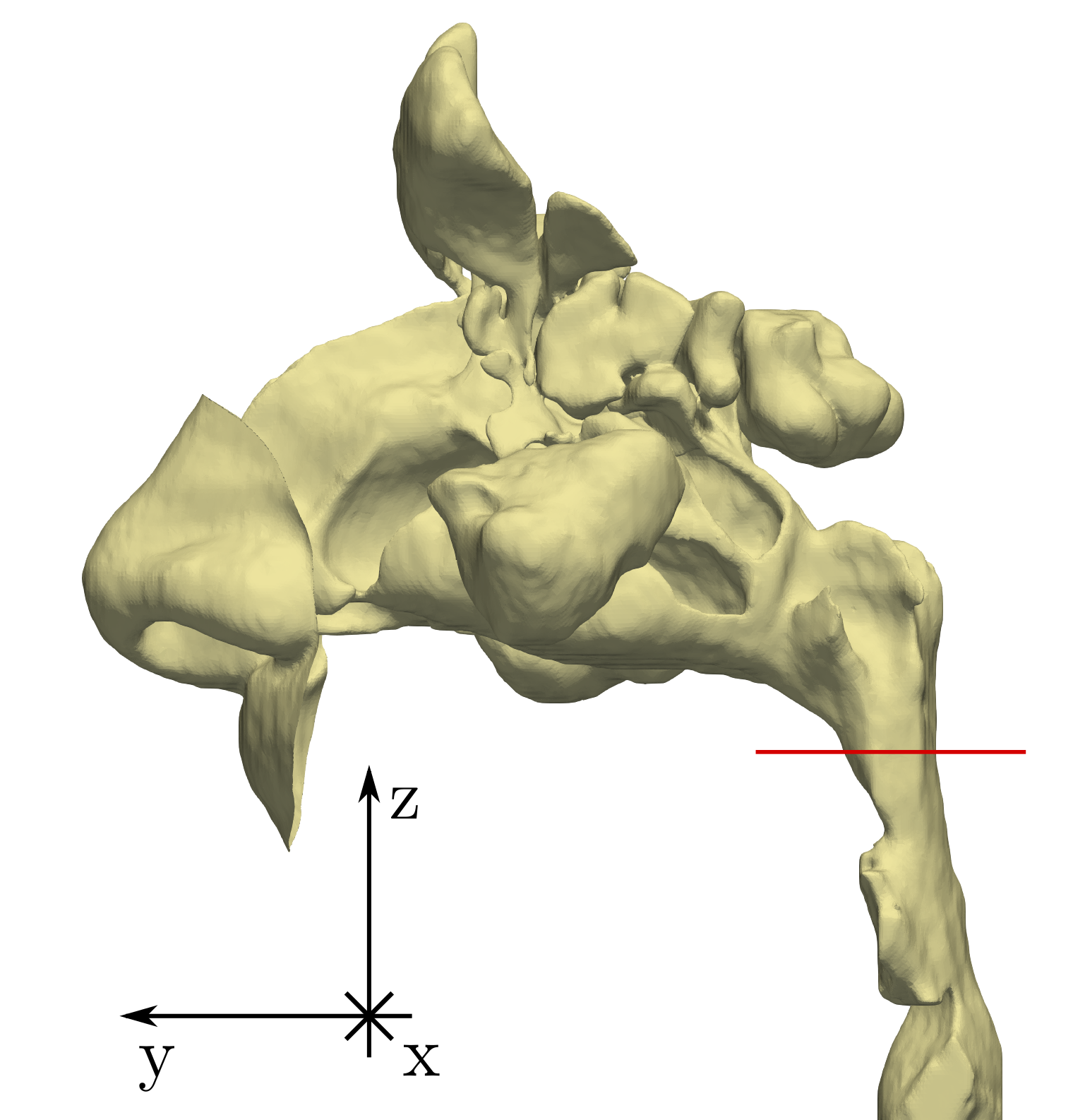}
\includegraphics[width=\columnwidth]{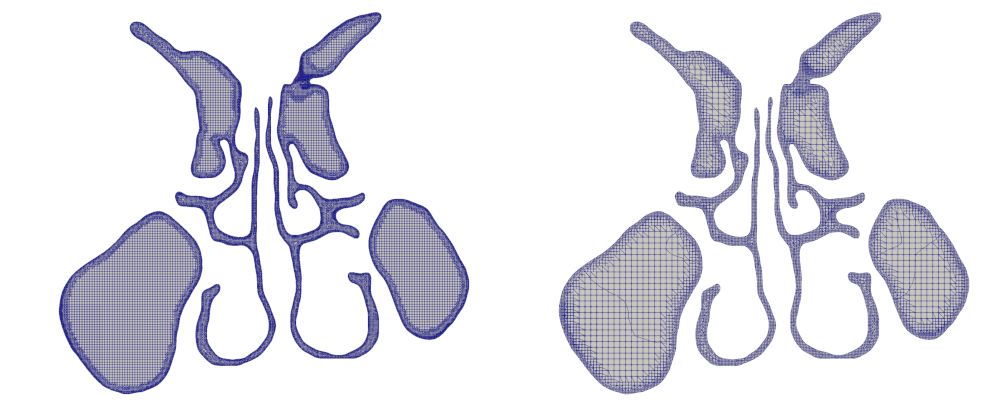}
\caption{Top: three-dimensional view of the CT reconstructed anatomy, the red line is where the volume is cut to mimic the TrCT anatomy. Bottom: coronal section of the volume mesh employed for LES (left) and RANS (right) simulations. Both feature a refinement near the solid boundary.}
\label{fig:mesh}
\end{figure}

This paper discusses results from 24 simulations, consisting in 12 inspiration and expiration pairs where every combination of i) first- and second-order numerical schemes, and ii) laminar, RANS and LES modeling is considered. The entire study is carried out twice, on standard (CT) and truncated (TrCT) volumes.  A larger LES case with second-order accuracy achieving quasi-DNS spatial resolution provides reference (inspiration only). A detailed comparison between CT and TrCT is described in the Supplementary Material, where additional details of the entire procedure are also mentioned. The various cases are indicated in this paper as for example CT-RANS-II-i, meaning CT-type scan, RANS modeling, second-order schemes, and inspiration. HRLES-II-i indicates the High-Resolution LES case. Normal breathing at rest is simulated by enforcing a steady volumetric flow rate of 280 $ml/s$ for all cases \citep[see e.g.][]{wang-lee-gordon-2012}. The baseline head CT scan is that of a male patient with healthy sinonasal anatomy. Figure \ref{fig:mesh} (top) presents the anatomy, reconstructed via standard CT segmentation procedures \citep{quadrio-etal-2015}, and also indicates where the original CT model is truncated above the epiglottis to obtain the TrCT version; the reference system used in the following is shown.

All simulations are incompressible and carried out within the OpenFOAM \citep{weller-etal-1998} finite-volumes software package, also used to create the volume mesh. The surface of the nasal cavities is considered as a solid wall, where no-slip and no-penetration boundary conditions are applied; pressure is set to zero at the outlet. The external ambient is represented via a sphere placed in front of the nose. RANS and LES require different meshes, and we have chosen their sizes to be typical of either approach, as determined from a broad literature scan: the RANS mesh has $3.2 \times 10^6$ cells (which drop to $2.8 \times 10^6$ for TrCT where the total volume is smaller) whereas the LES mesh has about $1.5 \times 10^7$ millions of cells ($1.4 \times 10^7$ for TrCT and  more than 50 millions cells for the reference HRLES). A mesh refinement analysis carried out for the RANS mesh and described in the Supplementary Material confirms its adequacy at properly describing the geometry and producing mesh-independent results. The flow is always solved down to the wall, and the use of wall functions is avoided. Figure \ref{fig:mesh} shows a comparison between the RANS and LES meshes. 


The RANS turbulence model is the $k-\omega-SST$ model, which is quite popular in such low-Reynolds and transitional flow, and was shown by \cite{li-etal-2017} to provide satisfactory results. The LES turbulence model is WALE (Wall-Adapting Local Eddy viscosity), which suits complex geometries well \citep{nicoud-ducros-1999}; the high spatial resolution makes the details of the LES model relatively unimportant. 


%% file: results.tex
\section{Results}
\label{results}

\begin{figure}
\includegraphics[width=\columnwidth]{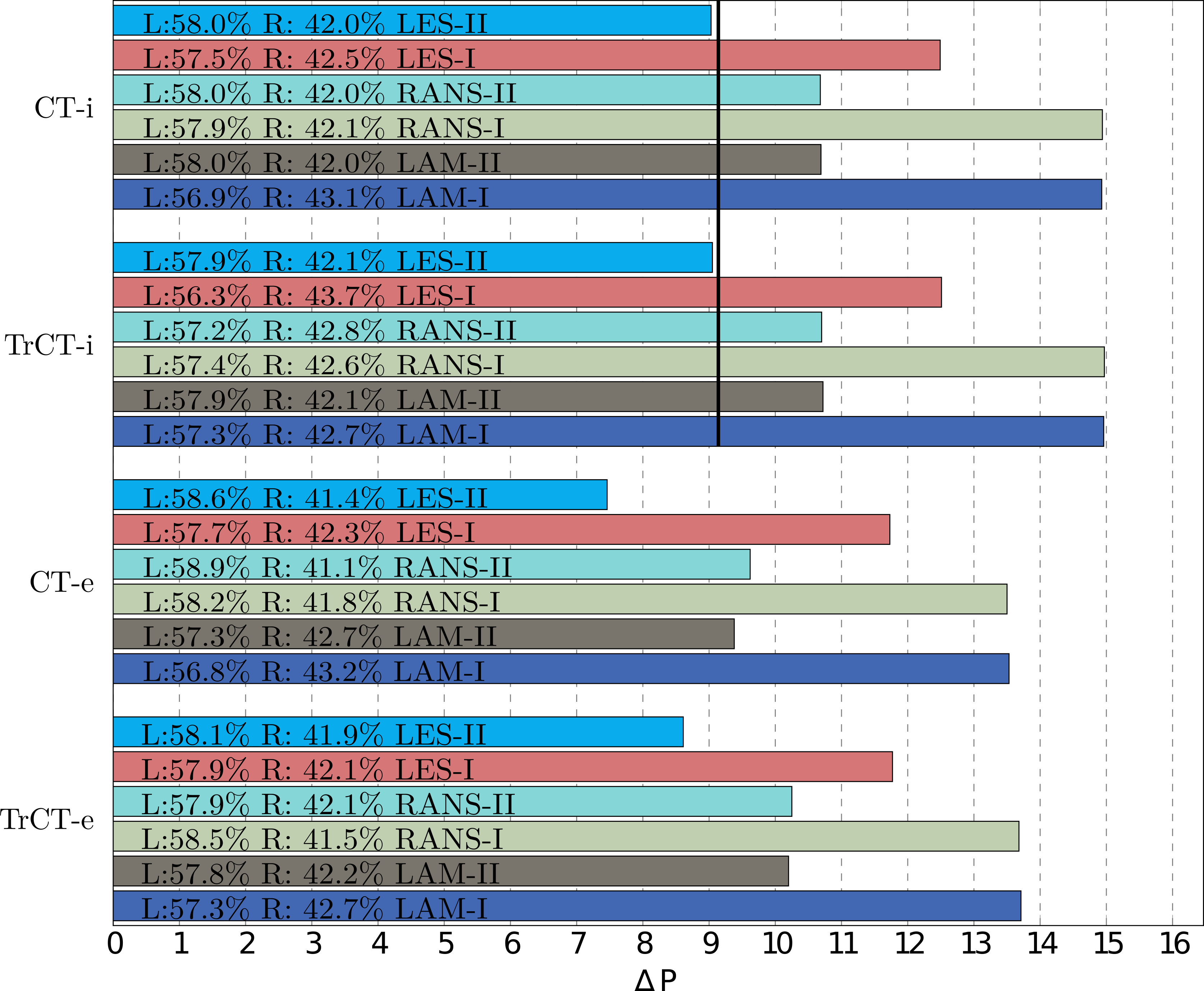}
\caption{Mean pressure difference $\Delta P$ between inlet and outlet, for all the computed cases. The percentage share of the flow rate in the left (L) and right (R) fossa is also shown within each bar. For CT cases, the measurement is taken at the red line shown in figure \ref{fig:mesh}. The vertical line is the reference pressure difference measured by HRLES-II-i.}
\label{fig:bars}
\end{figure}

The 24 cases are first compared in figure \ref{fig:bars} in terms of a global quantity, i.e. the (absolute value of the) mean pressure drop $\Delta P$ between the outer ambient and the lower end of the TrCT scan, marked by the red line in figure \ref{fig:mesh}. The percentage flow distribution in the left/right passageway is also displayed. Switching from first- to second-order schemes consistently reduces the pressure drop by about 4 $Pa$. RANS-I and LAM-I always predicts the highest pressure drop, followed by LES-I, RANS-II and LAM-II. LES-II, arguably the most reliable approach, provides the smallest pressure drop which is in agreement with HRLES-II. The left/right share of the flow is nearly unchanged, with about 58\% passing through the left and 42\% through the right, an asymmetry that \cite{borojeni-etal-2020} show to be well within normal values, in light of anatomical asymmetries and the effects of the nasal cycle. Switching from LAM/RANS to LES for the same numerical scheme brings the pressure drop down by about 1.5--2.5 $Pa$. 

\begin{figure*}
\centering
\includegraphics[width=\textwidth]{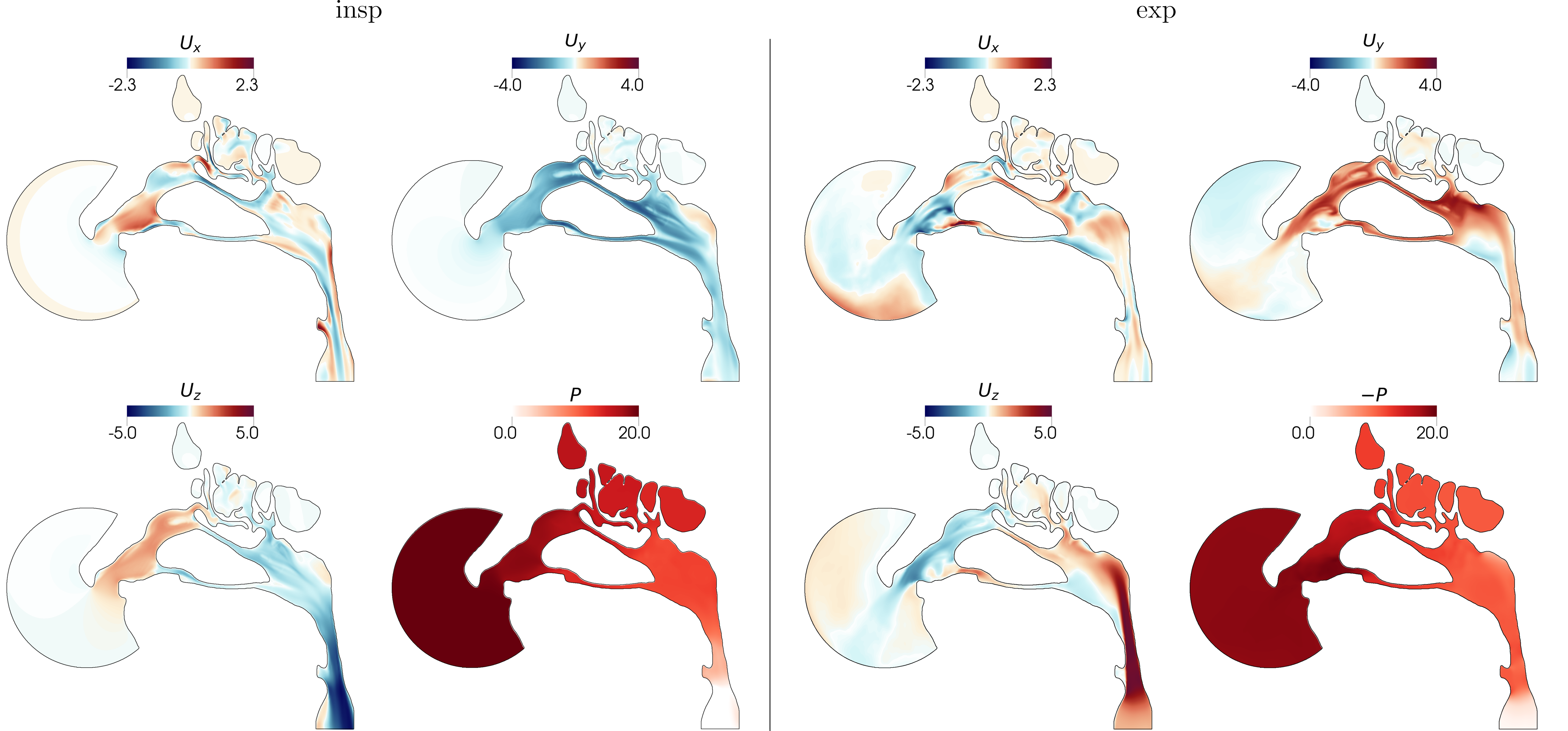}
\caption{Mean velocity and pressure fields in sagittal view. Left: CT-LES-II-i; right: CT-LES-II-e.}
\label{fig:insp_exp}
\end{figure*}

Before examining how these global changes reflect locally in the mean velocity and pressure fields, the general features of the solution (which is qualitatively similar across all cases) are briefly described. The mean fields computed in the CT-LES-II case are taken as example and shown in figure \ref{fig:insp_exp}. During the inspiration phase, the outer air is accelerated at the nostrils and then around the turbinates through the meati, with the velocity magnitude reaching up to 2--3 $m/s$. In the nasopharynx, the flow rotates downwards, but also produces a recirculation (visualized by the positive $U_y$ component) at the posterior wall of the nasopharynx. The largest velocity values in the flow field reach up to 4--5 $m/s$: this happens in particular for the $U_z$ component near the laryngeal stricture. Pressure, which is relative to the level $P=0$ set at the outlet, undergoes the largest drop under the epiglottis, in the lower region of the oropharynx. 

During expiration, air flows through a contraction at the laryngopharynx and produces a strong vertical jet, which impacts on the rear portion of the nasopharynx, then turns horizontally to enter the fossae and eventually reaches the outer ambient. The largest component is again $U_z$, as shown in Figure \ref{fig:insp_exp} (right), with a maximum of about 5 $m/s$. Pressure distribution qualitatively resembles the inspiration plot (except the direction of gradients), with the strongest drops at the larynx and in the meati. 

Having illustrated the general features of the mean flow field, we can proceed now to illustrate the changes induced by the parameters of interest.


\subsection{First- vs second-order schemes}

\begin{figure*}
\centering
\includegraphics[width=\textwidth]{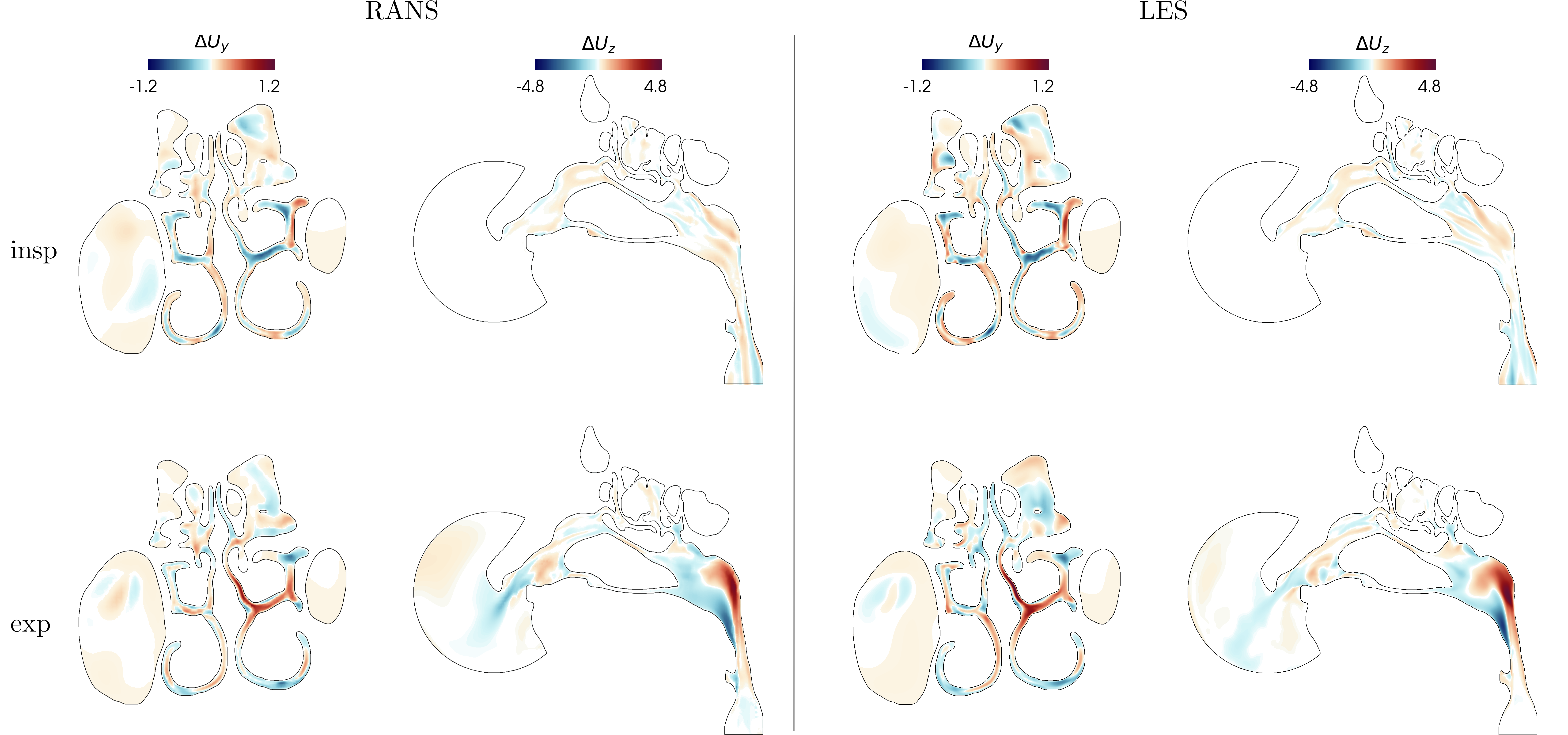}
\caption{Differential velocity field $\vect{U}_{II} - \vect{U}_I$: RANS (left) and LES (right) for the CT anatomy.}
\label{fig:NS}
\end{figure*}

Figure \ref{fig:NS} plots the two largest Cartesian components of the difference velocity field $\vect{U}_{II} - \vect{U}_I$, with $\vect{U}_I$ and $\vect{U}_{II}$ being the time-averaged velocity fields computed with first- and second-order schemes, respectively. 

In the RANS inspiration, differences up to 2.1 $m/s$ are found. In the coronal view, peak differences reside in the areas with the largest rate of flow, with maxima of 1.1 $m/s$ in the left inferior meatus and the right part of the middle meatus. The sagittal view shows significant velocity differences over the whole domain, except the external spherical volume and the sinuses. For the corresponding expiration, the coronal view shows similar differences still located in the middle meatus; the sagittal view, instead, shows a remarkable difference of 4.3 $m/s$ in the $U_z$ component, located in the nasopharynx. A rather similar picture is shown by the LES results, with comparable or even larger changes. To appreciate these differences, we observe that the bulk (area-averaged) velocity computed at the nostrils is 0.96 $m/s$.

\begin{figure}
\centering
\includegraphics[width=\columnwidth]{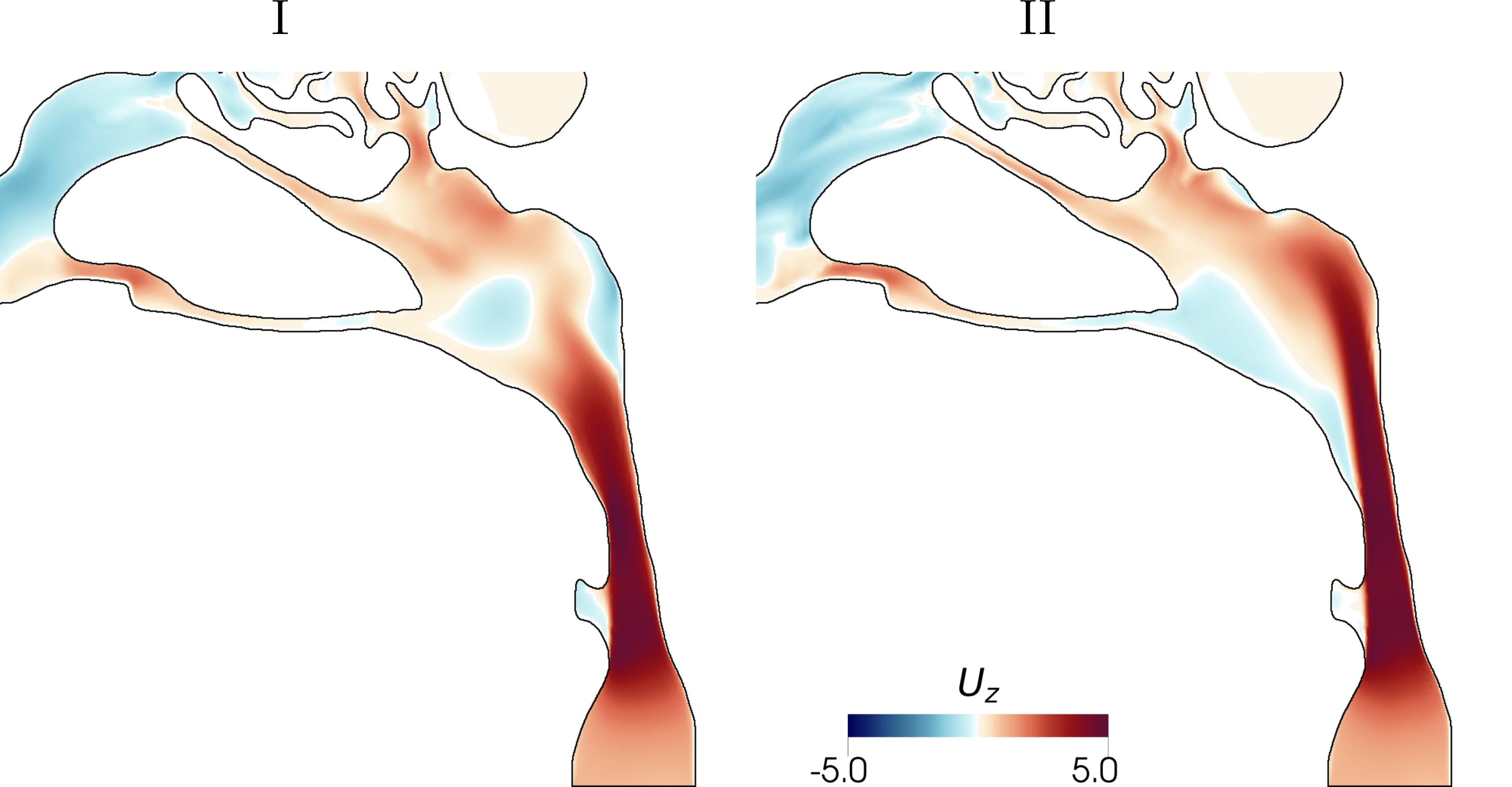}
\caption{Sagittal view of CT-LES-e: $U_z$ computed with first-order (left) and second-order (right) schemes.}
\label{fig:laryngeal-jet}
\end{figure}

Figure \ref{fig:laryngeal-jet} focuses on the largest changes, occurring in the laryngeal jet, and compares its spatial structure in expiration for numerical schemes of different accuracy. (Only LES is shown, RANS is similar.) The laryngeal jet is substantially different: the lower-accuracy case shows a rather short jet that ends within the nasopharynx, whereas the higher-accuracy case presents a longer, more coherent jet that crosses the entire pharynx and impacts on the posterior wall.

\subsection{RANS vs LES}

\begin{figure*}
\includegraphics[width=\textwidth]{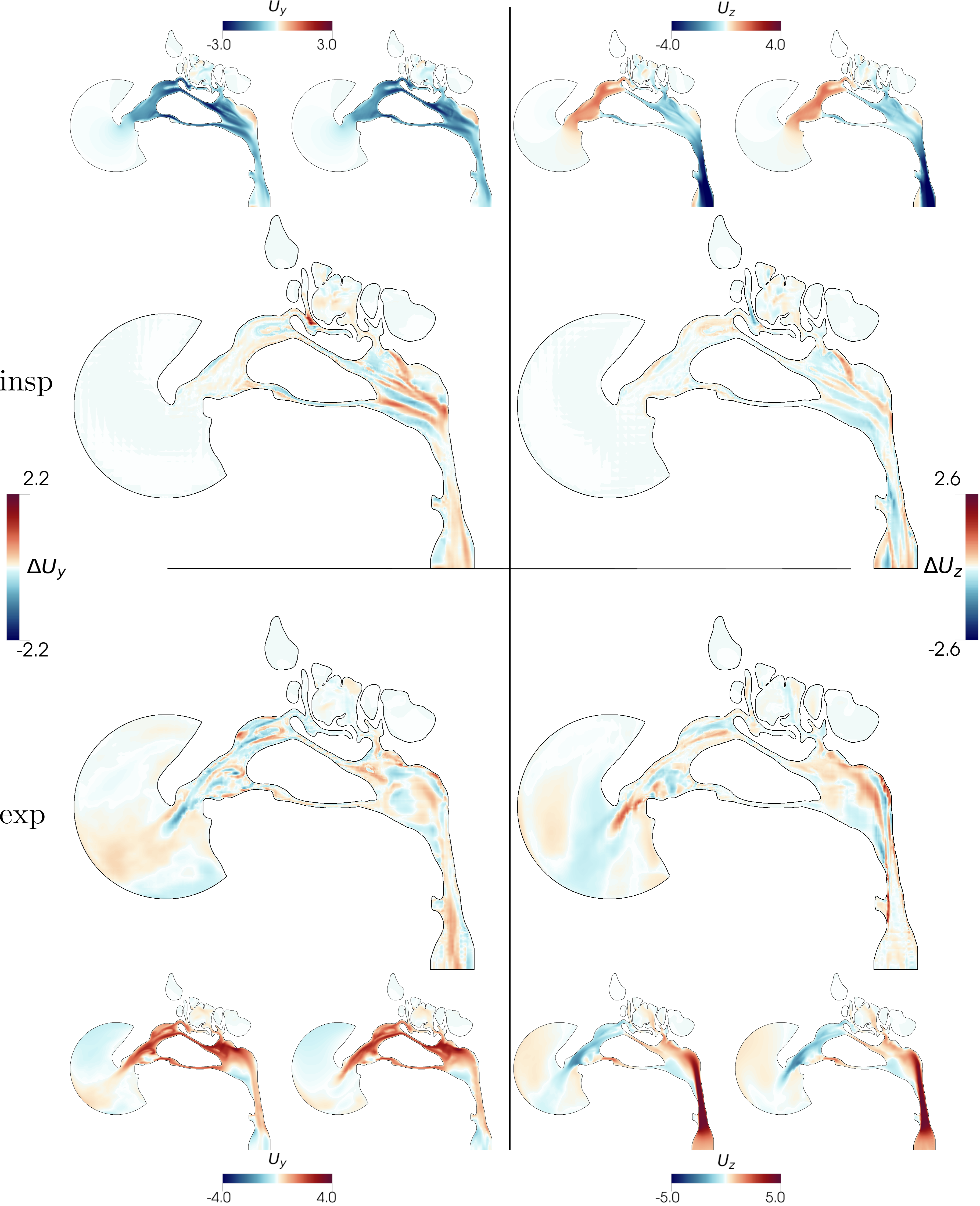}
\caption{Differential velocity field $\vect{U}_{LES} - \vect{U}_{RANS}$, for CT-II cases. The left and right columns describe the $U_y$ and $U_z$ velocity components respectively, while the top and bottom rows concern inspiration and expiration. For each panel, the largest figure plots the difference field, while the smallest panels plot the LES (left) and RANS (right) fields from which the difference field is generated.}
\label{fig:LR}
\end{figure*}

RANS and LES results are compared via the difference of their mean velocity fields, i.e. $\vect{U}_{LES} - \vect{U}_{RANS}$. Since these differences are found to be rather independent from the numerical scheme, only cases computed at second-order accuracy are shown in figure \ref{fig:LR}. The horizontal component $\Delta U_y$ reaches up to 2.2 $m/s$ in the area of the nasopharynx. In inspiration, differences are related to the shear layers detaching from the vestibular region; in expiration, differences extend to the meati. Especially during expiration, significant differences are observed in the vestibular area of the nose, of the order of 2 $m/s$ for both velocity components.

\begin{figure*}
\includegraphics[width=\textwidth]{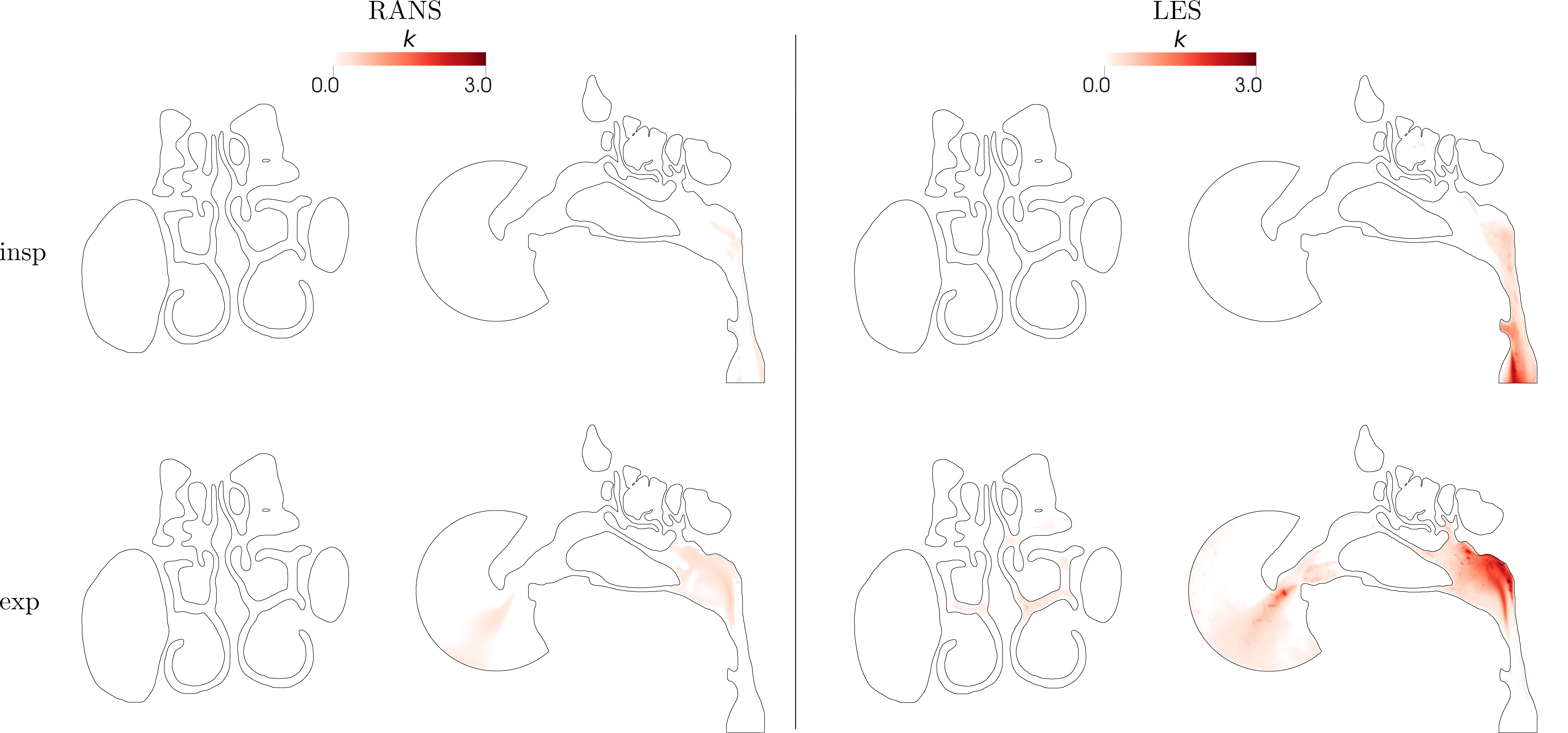}
\caption{Field of turbulent kinetic energy $k$ as computed from CT-RANS-II (left) and CT-LES-II (right).}
\label{fig:k}
\end{figure*}

Significant differences are also expected in the correct representation of turbulence, and in particular the field of turbulent kinetic energy $k$, which is entirely modelled by RANS and computed by LES. Figure \ref{fig:k} confirms that $k$ largely differs between RANS and LES. 

%% file: discussion.tex
\section{Discussion}
\label{discussion}

The present results describe how the discretization scheme affects the CFD-computed airflow in the human nose, both globally and locally, and compares this effect to the modeling approach and to the type of CT scan.

The global effect has been quantified by measuring the pressure drop for a given flow rate. From figure \ref{fig:bars}, it appears that the formal order of accuracy of the discretization schemes plays a crucial role, independently from the flow model. On a given mesh, low-order numerical schemes are found to predict larger pressure drops, consistently with their more dissipative nature. Similarly, for a given numerical scheme, RANS predicts a larger pressure drop than LES, again because of the dissipative nature of the RANS turbulence models based on the concept of turbulent viscosity \citep{pope-2000}. The changes are substantial: at this flow rate, the pressure drops computed by a first-order RANS and by a second-order LES differ up to 6 $Pa$, which in the TrCT case is a difference of more than 60\%. Higher-order schemes imply a larger computational cost, but marginally so: we have measured a modest 15\% increase in CPU time for all the considered flow models. The large effect of the numerical scheme of choice is an important element to consider in the ongoing discussion, see e.g. \cite{cherobin-etal-2020} and \cite{berger-etal-2021}, whether nasal resistance computed via CFD agrees with nasal resistance clinically measured with a rhinomanometer, and clearly advocates the specification of the employed numerical schemes in papers dealing with airflow in the human nose: the overestimate of the pressure drop by lower-accuracy methods would further increase the gap between the two measuring techniques, while the scatter among CFD datapoints would be most certainly reduced. Unfortunately, however, in the current literature this essential information is not reported very often. 

Global differences arise as the integrated effect of a number of localized changes in the pressure and velocity fields. First-order numerical schemes misrepresent important parts of the flow physics, by for example failing to correctly capture the free shear layers in the nasopharynx during inspiration, or the massive laryngeal jet that develops during expiration. Use of CFD for detailed surgery planning would certainly benefit from a reliable representation of the whole flow physics, and thus mandates close attention to the numerical schemes employed in the CFD solution.

Flow modelling has been discussed multiple times in the past, and it comes at no surprise that laminar/RANS and LES outcomes are quite different, in terms of both pressure and velocity fields. Pressure differences indicate that RANS overestimate pressure drop by 2--4 $Pa$, independently from the numerical schemes; velocity differences are more delicate to interpret. The most affected flow region seems to be where free shear layers develop (the nasopharynx, and the vestibular area during expiration). Laminar/RANS modelling, although perhaps acceptable for normal sino-nasal anatomies like the present one, might become questionable once anatomic anomalies are present and disturb the flow field, leading to a more complex flow even in the relatively quiescent yet surgically delicate region of the nasal meati. Obviously, this has to be considered jointly with the different computational cost: speaking of CPU time alone, the typical mesh sizes used here lead to LES being approximately 60 times more expensive than RANS. Significant differences have been also found in the correct representation of turbulence, e.g. the turbulent kinetic energy field shown in figure \ref{fig:k}, thus reinforcing the case for the inadequacy of RANS modelling whenever anatomic anomalies induce significant localized flow unsteadiness.

This study has also considered the effect of a computational domain truncated well above the larynx, as it would happen when cone-beam CT scans are used. Changing the position of the lower boundary has little influence when inspiration is computed, but expiration is much more affected: the lack of the laryngeal restriction makes the laryngeal jet impossible to predict correctly. Given the undeniable convenience of cone-beam scans, and the importance of imparting lower radiation doses to the patient, we envisage the need for a suitable inlet boundary condition for expiration to implicitly compensate for the missing part of the domain.


Discussing differences between velocity fields would be incomplete without recalling that alternate ways exist to compare two vector fields. For example, one should be aware that looking at the Cartesian components of the velocity difference vector might misrepresent changes that would appear under different light if e.g. the modulus of the difference is considered. Also, differences should be evaluated by bearing in mind the intensity of the local mean value. 

\begin{figure}
\includegraphics[width=\columnwidth]{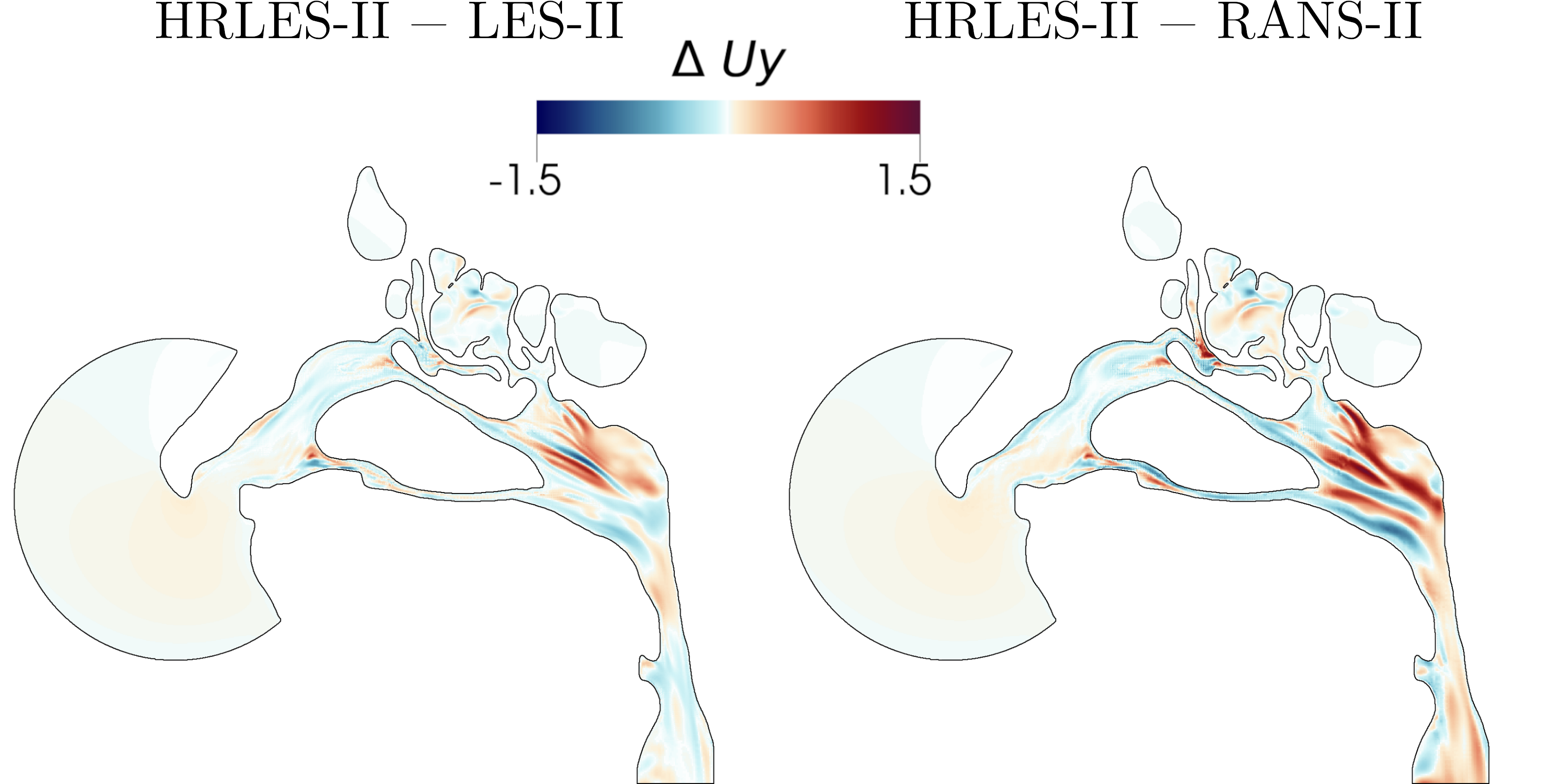}
\caption{Differential velocity field (sagittal component) HRLES-II - LES-II (left) and HRLES-II $-$ RANS-II (right).}
\label{fig:HRLES}
\end{figure}

Finally, so far we have discussed "differences" with the implicit assumption that LES-II naturally represents the most accurate approach in terms of both turbulence modelling versus RANS-II and numerics versus LES-I. However, LES-II results themselves are affected by modelling and discretization error: they would become error-free only on a very fine mesh. It is thus instructive to compare LES-II with the result of HRLES-II, where the larger mesh with 50 millions cells (more than 3 times the cells of LES-II) makes it approach the DNS limit. The global result of HRLES-II was already plotted as inspiration reference in figure \ref{fig:bars}; now figure \ref{fig:HRLES} clearly shows how LES-II is nearer than RANS-II to the reference, with residual errors that decrease both in spatial extension and absolute value as the spatial resolution increases and the LES modelling improves accordingly.